\documentclass[a4paper,oneside,12pt]{article}
\usepackage{amssymb,amsmath,amsfonts,amsthm,epsfig,multirow}
\usepackage{rotating}
\usepackage[left=2cm,right=2cm,top=2cm,bottom=2cm,bindingoffset=0cm]{geometry}
\theoremstyle{definition}

\begin{document}
\normalsize
\setcounter{page}{1}%
\begin{center} \textbf{\uppercase{All-Pairs Shortest Paths Algorithm \\ for High-dimensional Sparse Graphs}}
\end{center}

\begin{center}{Urakov\,A.\,R., Timeryaev\,T.\,V.}
\end{center}

\begin{abstract}{Here the All-pairs shortest path problem on weighted undirected sparse graphs is being considered. For the problem considered, we propose ``disassembly and assembly of a graph'' algorithm which uses a solution of the problem on a small-dimensional graph to obtain the solution for the given graph. The proposed algorithm has been compared to one of the fastest classic algorithms on data from an open public source.}

\end{abstract}
\section*{Introduction}
\hspace*{\parindent} The APSP (all-pairs shortest path problem) is one of the most popular tasks in graph theory because the shortest paths between all pairs of vertices are used for solving many problems involving discrete optimization (TSP, theory of transportation task etc). Moreover, the task itself is of great interest in research.

Recently this problem has gained new interest due to a growing number of highly detailed graphs that are generated automatically and describe structures from the real world. Such graphs have about $10^6$ or more vertices and this number will inevitably increase. So the acceleration of APSP solving for high-dimensional graphs is becoming highly important.

Because of its popularity, there are a lot of APSP solution algorithms but there isn't any method to obtain the solution as fast for different kinds of input data. That's why APSP solution algorithms can be classified according to the type of graph as follows: directed \cite{bellman58}, complete \cite{floyd62}, weighted \cite{dijkstra59}, unweighted \cite{cormen06} and sparse \cite{johnson77}.

Here we present an algorithm for solving the APSP for weighted, undirected and high-dimensional sparse graphs with non-negative weights.

This paper is organized as follows. In section \ref{notationSection}, we introduce notation and the problem definition, in section \ref{algorithmSection}, we describe the algorithm and in section \ref{resultsSection} we show the results in comparison with one of the most renowned APSP algorithms.

\label{notationSection}\section{Notation and problem definition}
\subsection{Terms and definitions}

\hspace*{\parindent} Here, we consider a connected, undirected and sparse graph $G=\left(V,E,w\right)$, where each edge $e\left(v_i,v_j\right)$ has a non-negative weight $w\left(i,j\right)$. The given graph $G$ is considered to be simple (has no loops or multiple edges).

Denote by $|V|=n$ \emph{the order of a graph} or cardinality of vertices set. Denote by $|E|=m$ \emph{the size of a graph} or cardinality of edges set.

Denote by $w\left(i,j\right)$ \emph{the weight} of the edge between vertices $v_i$ and $v_j$ ($w\left(i,j\right)=\infty$, for non-connected vertices). \emph{A degree} $d\left(v_i\right)$ of vertex $v_i$ is the number of edges incidental to $v_i$. A graph is called \emph{sparse} if $m\ll n^2$.

\emph{A path} is an alternating sequence of vertices and edges $v_0$, $e_1$, $v_1$, $\dots$, $v_{k-1}$, $e_k$, $v_k$, beginning and ending with a vertex. In that sequence, each vertex is incidental to both the edge that precedes it and the edge that follows it. \emph{A length} of a path is the sum of the weights of its edges. \emph{A distance} $m\left(i,j\right)$ between $v_i$ and $v_j$ is the length of the shortest path $p_{ij}^s=p^s(v_i,v_j)$ between these vertices. \emph{A distance matrix} is a matrix in which each element at the intersection of $i$th row and $j$th column contains the length of the shortest path between $v_i$ and $v_j$. A graph is said to be \emph{connected} if every pair of vertices in the graph is connected by some path, i.e. $m_{ij}<\infty,~\forall i,j$.

Between any pair of vertices there can be more than one shortest path. We do not consider it as an essential issue in this paper, so the references to the shortest path can mean any of them.

A matrix is called \emph{a precedence matrix} if each element $p_{ij}$ of the matrix corresponds to the vertex that precedes vertex $v_j$ in the path from $v_i$ to $v_j$. Therefore the elements of $P$ can be determined by
\begin{equation*}
p_{ij}=\begin{cases}
    v_k,&\exists v_k:p_{ij}^s=\dots v_k,e\left(v_k,v_j\right),v_j\\
    \infty,&\text{else}
    \end{cases}
\end{equation*}
Using $P$ the shortest path $p_{ij}^s$  from $v_i$ to $v_j$ in a connected graph can be obtained by the recursive formula:
\begin{equation*}
p_{ij}^s=\begin{cases}
     p^s\left(v_i,p_{ij}\right),e\left(p_{ij},v_j\right),v_j,&p_{ij}\neq v_i\\
     v_i,e\left(v_i,v_j\right),v_j,&p_{ij}=v_i
     \end{cases}
\end{equation*}

Now, we shall give the following supplementary definitions. Let us call a graph sequence $S=\left\{G_1,G_2,\dots,G_r\right\}$ \emph{shrinking} graph $G_0=\left(V_0,E_0,w_0\right)$, where $G_p=\left(V_p,E_p,w_p\right)$, $V_p=\left\{v_1^p,v_2^p,\dots,v_{n(p)}^p\right\}$, $E_p=\left\{e_1^p,e_2^p,\dots,e_{m(p)}^p\right\}:e_i^p=e^p\left(v_j^p,v_k^p\right)\subseteq V_p \times V_p$
and $w_p:E_p\rightarrow [0,\infty)$.

Every next graph $G_{p+1}$ of the sequence is obtained from the previous $G_p$ by the removing the $k$ vertices and the edges incidental to them, plus the addition of new edges and by recalculating the weights of the edges adjacent to the deleted ones.

For these graphs, we get $\left|V_p\right|>\left|V_{p+1}\right|,\forall p=\overline{0,r-1}$.
Denote by $v_i^{p+1}$ a vertex of $G_{p+1}$ corresponding to vertex $v_i^p$ of $G_p$.
Denote by $e^{p+1}\left(v_{j}^{p+1},v_{k}^{p+1}\right)$ an edge of $G_{p+1}$ corresponding to the edge $e^p\left(v_j^p,v_k^p\right)$ of $G_p$.

Denote by $G_{p+1}=R_p\left(v_1^p,v_2^p,\dots,v_k^p\right)$ the graph obtained from $G_p$ by removing the vertices $v_1^p,v_2^p,\dots,v_k^p$ and the edges incidental to them. For this graph we get $w_{p+1}(i,j)=w_p(i,j),\forall i,j:v_i^{p+1},v_j^{p+1}\in V_{p+1}$.

Denote by $m^p(i,j)$ the distance between $v_i^p$ and $v_j^p$  in $G_p$. By $M_p=\left(m_{ij}^p\right)$ denote the distance matrix of $G_p$.
Also denote by $v_{i_l}^p$ $l$th adjacent to $v_i^p$ vertex and by $A_i^p$ the set of all adjacent to $v_i^p$ vertices in graph $G_p$.

\subsection{Problem definition}
% \textbf{Задача.}
\hspace*{\parindent} Given a connected, undirected, simple, weighted and sparse graph $G=(V,E,w)$, where each edge has a non-negative weight $w:E\rightarrow [0,\infty)$. Find the shortest paths between every pair of vertices of the graph, i.e. find the distance matrix $M$ and the precedence matrix $P$ of the graph.

\section{Algorithm of the solution}\label{algorithmSection}
\subsection{Main idea}
The main idea of the introduced algorithm is to reduce the problem on a large graph to the problem on a smaller graph. The algorithm can be partitioned into 3 stages.

\begin{enumerate}
 \item Compression. A large initial graph is replaced by a small graph.
 \item Microsolution. The APSP for the small graph is solved by using any known method.
 \item Restoring. The APSP solution for the small graph is projected onto the initial graph.
\end{enumerate}

While using this method we must satisfy the following conditions: a) validity "---the compression must keep information about the shortest paths of the initial graph; b) efficiency "---the introduced method must be quicker than all others.

The algorithm in which similar ideas were used are considered in \cite{geisberger08}.
Here we introduce an algorithm of a graph disassembly/assembly for large sparse graphs. At the disassembly stage, we consistently remove vertices, and then solve the APSP for the resulting small graph. At the assembly stage the initial graph is restored with the calculation of distances and paths.

\subsection{Disassembly}
\hspace*{\parindent} The disassembly stage consists of consistent approximation of the initial graph $G_0=\left(V_0,E_0,w_0\right)$ by the graphs of a shrinking sequence $S=\left\{G_1,G_2,\dots,G_r\right\}$. Here we consider a particular case in which every next graph $G_{p+1}$ of the sequence $S$ is obtained by removing only one vertex from $G_p$.

Suppose that vertex $v_i^p$ is to be removed. Let the degree of $v_i^p$ be equal to $k$. If any shortest path contains $v_i^p$ (except shortest path straight to or from $v_i^p$) then this path contains subpath $v_{i_j}^p,e^p(v_{i_j}^p,v_i^p),v_i^p,e^p(v_i^p,v_{i_l}^p),v_{i_l}^p:j,l\in \{1,2,\dots,k\}$. Therefore to remove vertex $v_i^p$ properly, we need to preserve the shortest paths only between vertices adjacent to $v_i^p$.

By $w_p^{\text{mv}(1,2,\dots,h)}\left(i_j,i_l\right)=\min_{g=1,2,\dots,h}\left(w_p\left(i_j,g\right)+w_p\left(g,i_l\right)\right)$ denote the minimum sum of the weights of two edges which connect vertices $v_{i_j}^p$, $v_{i_l}^p$ and are incidental to a common vertex that belongs to the set $v_1^p,v_2^p,\dots,v_h^p$ of $G_p$. To preserve distances it is sufficient to have
\begin{equation}\label{eq:exp1}
 w_{p+1}\left(i_j,i_l\right)=\begin{cases}
                           \min\left( w_p^{\text{mv}(i)}\left(i_j,i_l\right),w_p\left(i_j,i_l\right)\right),&\text{}w_p^{\text{mv}(i)}\left(i_j,i_l\right)<w_p^{\text{mv}(h\neq i)}\left(i_j,i_l\right) \\
                           w_p\left(i_j,i_l\right),&\text{else}
                          \end{cases}
\end{equation}
for any pair $\left(v_{i_j}^p,v_{i_l}^p\right)$ in $G_{p+1}$.

At the beginning of the algorithm any element of $P^{'}$ is equal to infinity $p^{'}_{ij}=\infty,~\forall i,j$. To preserve the information about the shortest paths, for each element of $P^{'}$ that satisfies $w_p^{\text{mv}(i)}\left(i_j,i_l\right)<\min\left(w_p\left(i_j,i_l\right),w_p^{\text{mv}(h\neq i)}\left(i_j,i_l\right)\right)$ we have
\begin{equation}\label{eq:exp2}
 p^{'}_{i_ji_l}=\begin{cases}
         v_i,&\text{} p^{'}_{ii_l}=\infty\\
         p^{'}_{ii_l},&p^{'}_{ii_l}\neq\infty
        \end{cases}
\end{equation}

Note: if vertex $v_i^p$, which is to be removed, is adjacent only to one vertex of $G_p$, so, as there are no shortest paths passing through $v_i^p$, the vertex and the incidental edge are simply removed without the shortest path preservation.

We use three parameters for the disassembly stage. $d_{max}$ "---is the maximum degree of the vertices to be removed. $n_{min}$ "---is the order of graph $G_r$, which is the last (smallest) graph of the shrinking sequence. $I_{max}$ "---is the limit of the increasing number of edges after the removal of one vertex. The assignment of values to $d_{max}$, $n_{min}$ and $I_{max}$ is a problem in itself, which will be discussed elsewhere. The results, which are shown in part \ref{resultsSection}, have been obtained by assignment $d_{max}=I_{max}=\infty$, $n_{min}=1$.

Let us try to remove vertex $v_i^p$ with all of its $k$ incidental edges and preserve the shortest paths. Denote by $I\left(v_i^p\right)$ the change in the number of graph edges when the vertex is removed. The removal of $v_i^p$ itself will decrease the number of edges by $k$, therefore we get $I\left(v_i^p\right)=-k$. Using the shortest paths preservation and \eqref{eq:exp1}, we have:

\begin{equation}\label{eq:exp3}
 I\left(v_i^p\right)=\begin{cases}
         I\left(v_i^p\right)+1,&\text{if }w_p\left(i_j,i_l\right)=\infty~\wedge~w_p^{\text{mv}(i)}\left(i_j,i_l\right)<w_p^{\text{mv}(h\neq i)}\left(i_j,i_l\right)\\
         I\left(v_i^p\right),&\text{else}
        \end{cases}
\end{equation}

Thus we'll obtain the change in the size of graph $G_{p+1}$ relative to $G_p$ after the removal of  vertex $v_i^p$. If $I\left(v_i^p\right)>0$, then the graph size increases, otherwise the graph size decreases or remains the same.  Using \eqref{eq:exp3} we expect that the increase of the graph size is bounded above by $I_{max}$ when a vertex is removed. It follows that vertex $v_i^p$ can be removed only if $I\left(v_i^p\right)\leq I_{max}$.

The selection of the vertices that we are going to remove is performed in the following way. Since vertices meeting $d\left(v_i^p\right)<3$ can be removed anyway, it follows that vertices should be removed in ascending order of their degrees from 1 to $d_{max}$. This speeds up the algorithm due to a smaller number of processed vertices with degrees close to $d_{max}$. After we remove $v_i^p$, the degrees of the adjacent vertices can change, hence, if we remove $v_i^p$, the vertices adjacent to $v_i^p$ should be processed through recursion. The graph disassembly algorithm and an auxiliary algorithm of vertices inspection and removal are on fig. 1 and 2.

% тут can удачнее could - типа "могли бы", "хотели бы"

\smallskip
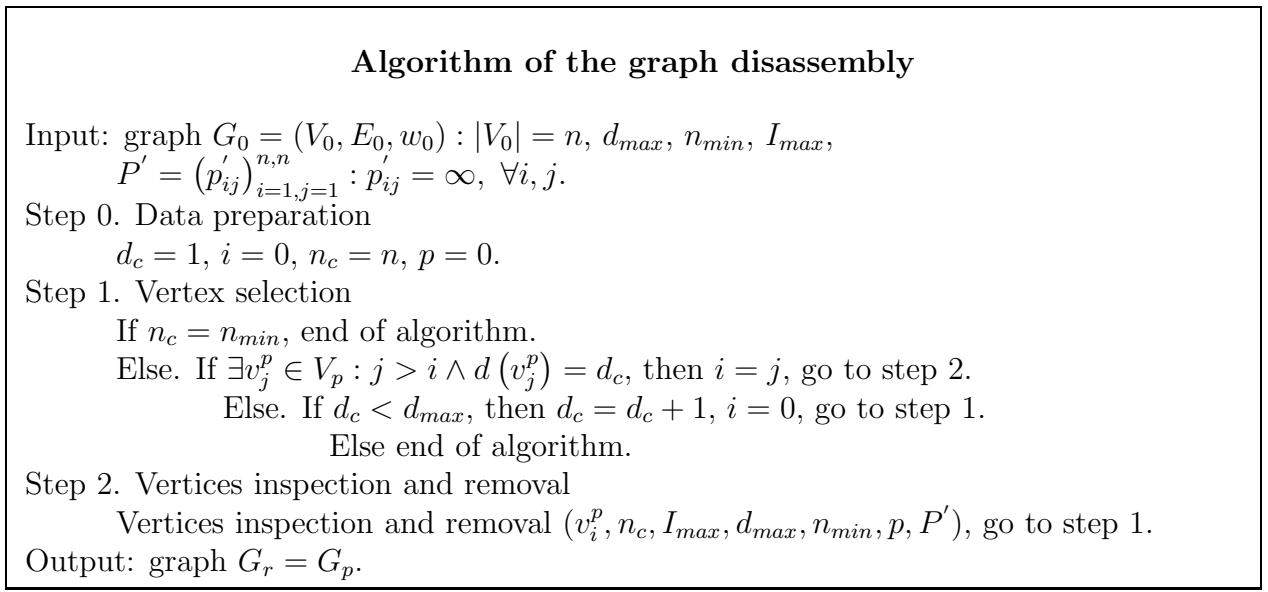
\begin{center}
\fbox{
\parbox{16cm}{
\begin{center} \textbf{Vertices inspection and removal}\end{center}
\hangindent=1.2cm \noindent Input: vertex $v_i^p$, number of vertices $n_c$, $I_{max}$, $d_{max}$, $n_{min}$, $p$, $P^{'}$.

Step 1. Vertices inspection

\hangindent=1.2cm \hangafter=-1 If $d\left(v_i^p\right)<3$, go to step 2.

\hangindent=1.2cm \hangafter=-1 Else $I\left(v_i^p\right)=-d\left(v_i^p\right)$. Inspect all pair of vertices $A_i^p$ and

\hangindent=2.6cm \hangafter=-1 change $I\left(v_i^p\right)$ by \eqref{eq:exp3}.

\hangindent=2.6cm \hangafter=-1 If $I\left(v_i^p\right)\leq I_{max}$, go to step 2.

\hangindent=2.6cm \hangafter=-1 Else end of algorithm.

\hangindent=\parindent Step 2. Vertex removal

\hangindent=1.2cm \hangafter=-1 Form a new graph $G_{p+1}=R_p\left(v_i^p\right)$.

\hangindent=1.2cm \hangafter=-1 Count the weights of the edges between vertices  $A_i^p$ by \eqref{eq:exp1}.

\hangindent=1.2cm \hangafter=-1 Change the elements of the matrix $P^{'}$ by \eqref{eq:exp2}.

\hangindent=1.2cm \hangafter=-1 $n_c=n_c-1$, $t=p$, $p=p+1$.

\hangindent=1.2cm \hangafter=-1 If $n_c=n_{min}$, end of algorithm.

\hangindent=1.2cm \hangafter=-1 Else, while $n_c>n_{min}$ for vertices $v_{i_l}^{p}:d\left(v_{i_l}^{p}\right)<d\left(v_{i_l}^t\right)$ do

\hangindent=2.6cm \hangafter=-1 Vertices inspection and removal ($v_{i_l}^{p},n_c,I_{max},d_{min},n_{min}, p, P^{'}$).
}
}
\end{center}

\begin{center}
 Fig. 1: Auxiliary algorithm of vertices inspection and removal.
\end{center}

\smallskip
\begin{center}
\fbox{
\parbox{16cm}{
\begin{center} \textbf{Algorithm of the graph disassembly}\end{center}
\hangindent=1.2cm \noindent Input: graph $G_0=(V_0,E_0,w_0):|V_0|=n$, $d_{max}$, $n_{min}$, $I_{max}$,

\hangindent=1.2cm \hangafter=-1 $P^{'}=\left(p^{'}_{ij}\right)_{i=1,j=1}^{n,n}:p^{'}_{ij}=\infty,~\forall i,j$.

\hangindent=\parindent Step 0. Data preparation

\hangindent=1.2cm \hangafter=-1 $d_c=1$, $i=0$, $n_c=n$, $p=0$.

\hangindent=\parindent Step 1. Vertex selection

\hangindent=1.2cm \hangafter=-1 If $n_c=n_{min}$, end of algorithm.

\hangindent=1.2cm \hangafter=-1 Else. If $\exists v_j^p\in V_p:j>i \wedge d\left(v_j^p\right)=d_c$, then $i=j$, go to step 2.

\hangindent=2.6cm \hangafter=-1 Else. If $d_c<d_{max}$, then $d_c=d_c+1$, $i=0$, go to step 1.

\hangindent=4cm \hangafter=-1 Else end of algorithm.

\hangindent=\parindent Step 2. Vertices inspection and removal

\hangindent=1.2cm \hangafter=-1 Vertices inspection and removal ($v_i^p,n_c, I_{max}, d_{max}, n_{min}, p, P^{'}$), go to step 1.

\hangindent=\parindent Output: graph $G_r=G_p$.
}
}
\end{center}

\begin{center}
 Fig. 2: Algorithm of the graph disassembly.
\end{center}

\subsection{Microsolution}
\hspace*{\parindent} Here the APSP for $G_r$ is solved. The result of the stage is the distance matrix $M_r$ of $G_r$.
We use matrix $M^{'}_r=M_r$ and recalculate $P^{'}$ by
\begin{equation}\label{eq:exp4}
 p^{'}_{ij}=\begin{cases}
                           p_{ij}^r,& p^{'}_{ij}=\infty~\wedge~p^{'}_{p_{ij}^rj}=\infty\\
                           p^{'}_{p_{ij}^rj},& p^{'}_{ij}=\infty~\wedge~p^{'}_{p_{ij}^rj}\neq\infty
                          \end{cases}
\end{equation}
\begin{equation}\label{eq:exp5}
 p^{'}_{ij}=\begin{cases}
                           p_{ij}^r,&w_r(i,j)>m^r_{ij}~\wedge~p^{'}_{p_{ij}^rj}=\infty\\
                           p^{'}_{p_{ij}^rj},&w_r(i,j)>m^r_{ij}~\wedge~p^{'}_{p_{ij}^rj}\neq\infty
                          \end{cases}
\end{equation}
here $p_{ij}^r$ are the elements of the matrix $P_r=\left(p_{ij}^r\right)$, which corresponds to $G_r$. The calculated paths are the shortest ones due to the usage of the distances preservation method. In other words, we have $m^{'r}_{ij}=m_{ij}^r=m^0_{ij},\forall i,j:v_i^r, v_j^r\in V_r$.

Obviously, if $G_r$ has only one vertex then this stage is skipped and the assembly of the graph starts.

\subsection{Assembly}
\hspace*{\parindent} Before this stage starts, the graph assembly sequence $S=\left\{G_0,G_1,\dots,G_r\right\}$ is defined. Here $G_0$ "---is the initial graph, $G_r$ "---is the smallest graph. The shortest paths between all vertices of $G_r$ were found in the previous stage. At the assembly stage we restore the removed vertices in reverse order to their removal. That is we move from $G_r$ to  $G_0$ through $G_{r-1},G_{r-2},\dots,G_1$, recalculating the shortest paths for vertex $v_{i}^{r-p}:v_{i}^{r-p+1}\notin V_{r-p+1}\wedge v_i^{r-p}\in V_{r-p}$ in each step $p$.

Suppose vertex $v_i^{r-1}$ is to be restored, i.e. we move from $G_r$ to $G_{r-1}$. Vertex $v_i^{r-1}$ is connected with vertices $\left\{v_{i_z}^{r-1}\right\}_{z=1}^k$ by $k$ edges. Matrix $M^{'}_r=M_r$ of $G_r$ was found in the previous step, therefore to find the matrix $M^{'}_{r-1}$ of $G_{r-1}$, we only need to calculate the shortest paths from vertex $v_i^{r-1}$ to all other vertices of $G_{r-1}$.  Other elements of $M^{'}_{r-1}$ are assigned equally to the corresponding elements of $M^{'}_r$, that is $m_{jl}^{'r-1}=m_{jl}^{'r},\forall j,l:v_j^r,v_l^r\in V_r$.

Since the shortest path from any vertex of $G_{r-1}$ to $v_i^{r-1}$ goes through $\left\{v_{i_z}^{r-1}\right\}_{z=1}^k$, it follows that the distance from $v_i^{r-1}$ to any vertex $v_l^{r-1}$ of $G_{r-1}$ can be calculated by $m^{'r-1}\left(i,l\right)=\min_{z=\overline{1,k}}\left(w_{r-1}\left(i,i_z\right)+m^{'r}\left(i_z,l\right)\right)$.

If we move from $G_{r-p+1}$ to $G_{r-p}$ by adding vertex $v_i^{r-p}$, to obtaining the distance of matrix $M^{'}_{r-p}$, we should use the following

\begin{equation}\label{eq:exp6}
 m_{jl}^{'r-p}= m_{jl}^{'r-p+1},~\forall j,l: v_j^{r-p+1},v_l^{r-p+1}\in V_{r-p+1}
\end{equation}

\begin{equation}\label{eq:exp7}
 m_{il}^{'r-p}=m_{li}^{'r-p}=
                 \min_{z=\overline{1,k}}\left(w_{r-p}\left(i,i_z\right)+m^{'r-p+1}\left(i_z,l\right)\right),~\forall l: v_l^{r-p+1}\in V_{r-p+1}
\end{equation}

Denote by $x(l)$ the number $i_z$ such that \begin{math}x(l):w_{r-p}\left(i,x(l)\right)+m^{'r-p+1}\left(x(l),l\right)= \\ \min_{z=\overline{1,k}}\left(w_{r-p}\left(i,i_z\right)m^{'r-p+1}\left(i_z,l\right)\right)\end{math}.
If any vertex satisfies $w_{r-p}(i,l)>m_{il}^{'r-p}~\vee~w_{r-p}(l,i)>m_{li}^{'r-p}$ or $p^{'}_{il}=\infty~\vee~p^{'}_{li}=\infty$, then the respective elements of matrix $P^{'}$ should be changed by

\begin{equation}\label{eq:exp8}
 p^{'}_{il}=\begin{cases}
                 v_i,&p^{'}_{x(l)l}=\infty\\%~\wedge~x(l)=l \\
                 %v_{x(l)},&p^{'}_{x(l)l}=\infty~\wedge~x(l)\neq l\\
                 p^{'}_{x(l)l},&p^{'}_{x(l)l}\neq\infty
        \end{cases}
\end{equation}

\begin{equation}\label{eq:exp9}
 p^{'}_{li}=\begin{cases}
                 v_l,&p^{'}_{x(l)i}=\infty\\%~\wedge~x(l)=i \\
                 %v_{x(l)},&p^{'}_{x(l)i}=\infty~\wedge~x(l)\neq i\\
                 p^{'}_{x(l)i},&p^{'}_{x(l)i}\neq\infty
        \end{cases}
\end{equation}

The assembly algorithm is shown in Figure 3.

\smallskip
\begin{center}
\fbox{
\parbox{16cm}{
\begin{center} \textbf{Assembly algorithm}\end{center}
\hangindent=1.2cm \noindent Input: $S=\left\{G_0,G_1,\dots,G_r\right\}$, $M^{'}_r$, $P^{'}$, $p=1$.

Step 1. Go to a larger graph

\hangindent=1.2cm \hangafter=-1 If $p\leq r$, count matrix $M^{'}_{r-p}$ by \eqref{eq:exp6},\eqref{eq:exp7} and matrix

\hangindent=2.6cm \hangafter=-1 $P^{'}$ by \eqref{eq:exp8} and \eqref{eq:exp9}, $p=p+1$.

\hangindent=1.2cm \hangafter=-1 Else end of algorithm.

\hangindent=\parindent Output: matrix $M^{'}_0$, matrix $P^{'}$.
}
}
\end{center}

\begin{center}
 Fig. 3: Assembly algorithm.
\end{center}

\section{Results}\label{resultsSection}
\hspace*{\parindent} All the tests have been performed on a computer equipped with an Intel Core 2 Duo E8400 (3 GHz) CPU and 2 GBs of RAM on the 32-bit edition of Windows XP. The source code has been written on C++ programming language in Borland C++ Builder 6. Weighted graphs of the USA road networks from an open public source ($G1-G10$) \cite{dimacsURL} have been used as the test data. Connected subgraphs with sizes from $10^3$ to $10^4$ of 10 pieces for each size have been derived from graphs $G1-G10$. Another set of test data are the graphs of Russian cities' road networks (($GR$, for detailed specifications look at \cite{urakov12}). The details of the test graphs are shown in \ref{tab1}.

\begin{table}[ht]
\centering
\caption{Characteristics of graphs used for testing}
\label{tab1}
\begin{tabular}{|c|c|c|c|c|c|}\hline
Group & Avg. quantity of & Avg. quantity of & Average & Max & \multirow{2}{*}{Graphs} \\
graphs & Vertices & Edges & vertex degree & vertex degree & \\ \hline
$G1$ & $10^3$ & $2,5\cdot10^3$ & 2,48 & 6 & \multirow{10}{*}{10} \\
\cline{1-5}
$G2$ & $2\cdot10^3$ & $5,21\cdot10^3$ & 2,6 & 5 & \\
\cline{1-5}
$G3$ & $3\cdot10^3$ & $7,88\cdot10^3$ & 2,62 & 6 & \\
\cline{1-5}
$G4$ & $4\cdot10^3$ & $1,07\cdot10^4$ & 2,68 & 6 & \\
\cline{1-5}
$G5$ & $5\cdot10^3$ & $1,33\cdot10^4$ & 2,66 & 6 & \\
\cline{1-5}
$G6$ & $6\cdot10^3$ & $1,58\cdot10^4$ & 2,63 & 7 & \\
\cline{1-5}
$G7$ & $7\cdot10^3$ & $1,85\cdot10^4$ & 2,64 & 6 & \\
\cline{1-5}
$G8$ & $8\cdot10^3$ & $2,08\cdot10^4$ & 2,6 & 6 & \\
\cline{1-5}
$G9$ & $9\cdot10^3$ & $2,36\cdot10^4$ & 2,62 & 7 & \\
\cline{1-5}
$G10$ & $10^4$ & $2,72\cdot10^4$ & 2,72 & 7 & \\
\hline
$GR$ & $2,1\cdot10^3$ & $6\cdot10^3$ & 2,86 & 14 & 20\\
\hline
\end{tabular}
\end{table}

Test parameters are $d_{max}=I_{max}=\infty,~n_{min}=1$. This means that all the graph vertices except one were deleted. That's why the microsolution stage was not performed. The proposed algorithm's (denote as \emph{PA}) has been compared to the binary heap implementation of the Dijkstra's algorithm (denote as \emph{DB}, \cite{cormen06}), which was performed for every vertex of the test graphs. The test results are shown in \ref{tab2}.

\begin{table}[ht]
\centering
\caption{Test results}
\label{tab2}
\begin{tabular}{|c|c|c|c|c|c|c|}\hline
\multirow{3}{*}{\begin{minipage}{1.3cm}\begin{center}Group\\of graphs\end{center}\end{minipage}} & \multirow{3}{*}{\begin{minipage}{1.95cm}\begin{center}PA avg.\\runtime, s\end{center}\end{minipage}} & \multirow{3}{*}{\begin{minipage}{1.95cm}\begin{center}DB avg. \\runtime, s\end{center}\end{minipage}} & \multirow{3}{*}{\begin{minipage}{1.8cm}\begin{center}PA max\\runtime, s\end{center}\end{minipage}} & \multirow{3}{*}{\begin{minipage}{1.8cm}\begin{center}DB max\\runtime, s\end{center}\end{minipage}} & \multirow{3}{*}{\begin{minipage}{2.4cm}\begin{center}Avg. speedup\\PA/DB\end{center}\end{minipage}} & PA max\\
& & & & & & deg. of rem.\\
& & & & & & vert.\\ \hline
$G1$ & 0,03 & 1,5 & 0,05 & 1,7 & 50 & 11 \\
\hline
$G2$ & 0,13 & 6,7 & 0,14 & 7,1 & 52 & 12 \\
\hline
$G3$ & 0,31 & 16 & 0,33 & 17 & 52 & 12 \\
\hline
$G4$ & 0,67 & 30 & 0,88 & 32 & 45 & 22 \\
\hline
$G5$ & 1,1 & 48 & 1,2 & 52 & 44 & 16 \\
\hline
$G6$ & 1,5 & 72 & 1,8 & 82 & 48 & 20 \\
\hline
$G7$ & 2,1 & 97 & 2,2 & 104 & 46 & 17 \\
\hline
$G8$ & 2,6 & 131 & 2,8 & 145 & 50 & 21 \\
\hline
$G9$ & 3,7 & 177 & 4,7 & 189 & 48 & 24 \\
\hline
$G10$ & 5,4 & 218 & 6,4 & 230 & 40 & 23 \\
\hline
$GR$ & 0,17 & 7,5 & 0,4 & 18,8 & 45 & 17 \\
\hline
\end{tabular}
\end{table}

The proposed algorithm speeds up the solving of APSP an average of 47 times faster in comparison with the Dijkstra algorithm. For each and all test graphs the algorithm is faster than the Dijkstra's algorithm (the minimum speed up is 34 times faster). During the tests, the vertices degrees were increased to a maximum of 17. This means that the complexity of the vertices removal increases during the disassembly only slightly.

\section*{Conclusion}
\hspace*{\parindent} The proposed algorithm noticeably accelerates the solving of the APSP for graphs of road networks, which is confirmed by the tests.
The objects of further research may be the selection of the algorithm parameters based on a fast analysis of graph properties, the modification of the disassembly and assembly order and the scalability issues of the algorithm relative to the increasing of a graphs' dimensions. Also, it is interesting to modify the algorithm to solve the problem quicker, but within a given error.


\begin{thebibliography}{8}
  \bibitem{cormen06}
    Cormen\,Т.\,H. at al.
    {\rm Introduction to algorithms. 2nd ed.\ MIT Press and McGraw-Hill, 2001}
  \bibitem{urakov12}
    Urakov\,A.\,R., Timeryaev\,T.\,V.
    {\rm Using weighted graphs features for fast searching their parameters // Prikl. Diskr. Mat. 2012. N\,2. C.\,95--99.}
  \bibitem{bellman58}
    Bellman\,R.
    {\rm On a routing problem // Quarterly of Applied Mathematics. 1958. N\,16. P.\,87--90.}
  \bibitem{dijkstra59}
    Dijkstra\,E.\,W.
    {\rm A note on two problems in connexion with graphs // Numerische Mathematik. 1959. N\,1. P.\,269--271.}
  \bibitem{floyd62}
    Floyd\,R.\,W.
    {\rm Algorithm 97: Shortest Path // Communications of the ACM. 1962. N\,5(6). P.\,345.}
  \bibitem{geisberger08}
    Geisberger\,R., Sanders\,P. at al.
    {\rm Contraction Hierarchies: Faster and Simpler Hierarchical Routing in Road Networks //    International Workshop on Experimental Algorithms (WEA 2008).}    Provincetown:~Springer, 2008. P.\,319--333.
  \bibitem{johnson77}
    Johnson\,D.\,B.
    {\rm Efficient algorithms for shortest paths in sparse graph // Journal of the ACM. 1977. N\,24. P.\,1--13.}
  \bibitem{dimacsURL}
    {\rm 9th DIMACS Implementation Challenge "---Shortest Paths. URL: http://www.dis.uniroma1.it/challenge9/download.shtml}
\end{thebibliography}
\end{document}